\documentclass[11pt]{article}
\usepackage{setspace}
\usepackage{amssymb}
\usepackage{amsmath}
\usepackage{amsfonts}
\usepackage{slashed}

\def\be{\begin{equation}}

\def\ee{\end{equation}}

\def\dd{\partial}

\def\bea{\begin{eqnarray}}

\def\eea{\end{eqnarray}}

\makeatletter
\def\blfootnote{\xdef\@thefnmark{}\@footnotetext}
\makeatother
\begin{document}

\singlespace

\begin{flushright} BRX TH-6647 \\
CALT-TH 2019-005
\end{flushright}

\vspace*{.3in}

\begin{center}

{\Large\bf Energy in Gravitation and Noether's Theorems}

{\large S.\ Deser}

{\it 
Walter Burke Institute for Theoretical Physics, \\
California Institute of Technology, Pasadena, CA 91125; \\
Physics Department,  Brandeis University, Waltham, MA 02454 \\
{\tt deser@brandeis.edu}
}
\end{center}

Dedicated to the memory of Peter Freund, friend and brilliant colleague for over five decades.

\begin{abstract}
I exhibit the conflicting roles of Noether's two great theorems in defining conserved quantities, especially Energy in General Relativity and its extensions: It is the breaking of coordinate invariance through boundary conditions that removes the barrier her second theorem otherwise poses to the applicability of her first. There is nothing new here, except the emphasis that General must be broken down to Special Relativity in a special, but physically natural, way in order for the Poincare or other global groups such as (A)dS to ``re-"emerge.
\end{abstract}

\section{Introduction}

The year 2018 sadly marks the end of Freund's life, a half-century since his very original work with Nambu (two masters of symmetry breaking) on (scalar) gravity [1] --- and, more happily, the Centennial of Emmy Noether's justly celebrated two theorems [2].  This induces the present pedagogical review of the notion of energy in General Relativity (and analogously of charge in nonabelian vector gauge models) to show that while its roots lie in both of Noether's results, it is defined only by sacrificing the second: breaking a local symmetry --- coordinate invariance --- is required. With Hilbert, Noether actually tried to apply her results to this question at the time, without --- for this reason --- succeeding [2]. That was to take another forty years [3]. After summarizing the theorems, I will turn to their applications in gauge theories and their sources, then to our main topic, their relevance to gravitation and its sources. All the facts are known; this gloss merely emphasizes the importance of gauge symmetry breaking in nonlinear models to define physical quantities there. Space limitations force omission of most essential details; they may be found in the references.

\section{Noether's Theorems}
Everyone knows them, so I merely remind their relevant form. The first says that invariance of an action under a continuous constant (also called global) transformation implies the existence of an accompanying conserved (on-shell) quantity (``charge"). The standard lore --- that general covariance forbids all (but scalar) constant invariances, hence that they are irrelevant --- is  true, but, as we will see, misleading.
The second theorem considers instead an action's invariance under local (``gauge") rotations, and links them to identical (off-shell) (covariant in general) conservation of the corresponding field equations. Put more tautologically, the gauge variation of a gauge invariant vanishes. 
Incidentally, Merton's first law --- no result is ever due to its namesake --- fails here: It's all Noether's!

\section{Gauge Theories}
As a relevant warm-up, we first consider vector gauge theories, which for our purpose are very different in their abelian, Maxwell, and non-abelian, Yang-Mills (YM) incarnations, because
the former are not charged, unlike YM excitations. Consider first source-free Maxwell; the second Noether theorem tells us merely that the field equations, $\dd_\beta F^{\beta\mu}$ are conserved identically. Indeed, this is true of any action involving $F^{\beta\mu}$, its dual and their derivatives: the field equations are of the manifestly conserved form $\dd_\beta H^{[\beta\mu]}=0$, where $H$ is an anti-symmetric tensor, also in 
arbitrary curved spaces. This has nothing to do with the invariance of any putative ``charged" sources, whose (as yet uncoupled to the photons) actions are invariant under constant rotations, and consequently specify a conserved charge, by the first theorem; here conservation is not identical but requires use of the field's equations. The four-current density $j^\mu$ is defined by introducing a fictitious coupling to a vector potential, varying with respect to the latter, then  setting it to zero. Of course Maxwell requires all its sources to be conserved to respect the second theorem, and the beauty of gauge-invariant couplings is that the matter $j^\mu$ remains on-shell conserved, if now dependent on  $A_\mu$ in general. [We often confuse our students by first extolling the beauty and essential role of gauge theories in physics, then tell them that gauge transformations are just an artifice to remove the extraneous components with which we have saddled the gauge theories to make them manifestly covariant, without mentioning the real reason --- locality!] So the abelian lesson its that Noether's two theorems are required separately for the left and right sides of the field equations.

Now comes the non-abelian difference: pure YM, like pure Maxwell, has a local invariance that implies identical conservation of its (now color) field equations, even though they read 
$(D_\beta H^{[\beta\mu]})^a =0$, where $D$ is the covariant derivative; $D_\mu$  and $H^{[\beta\mu]}$ both have color indices (denoted by $a$). While covariant conservation is still manifest because of the structure constants' anti-symmetry properties, it is very different from plain Maxwell-like conservation. The source story is now vastly different as well: YM excitations are ``charged", so while matter still has a covariantly conserved color current on-shell and both sides of the YM equations are again consistently conserved, by the two theorems, the notion of total charge gets complicated. It is essentially the same one as for GR energy, and indeed was only developed [4] after the latter was understood, so we only note it here. This is not to imply that the color ``charge" is entirely due to the sources: source-free YM solutions are charged and the charge is always due to the source+YM field, just as source-free gravity solutions have energy, as we will see.

Every gauge theory has as many constraint equations as invariances. These equations are of lower time derivative order that the rest (conservation itself is a ``$3+1$" notion!), and involve the 
corresponding time component(s) $j^0$ on the matter side. For Maxwell, this is the Poisson equation, whose space integral relates the longitudinal electric field to the conserved charge 
$Q= \int j^0 d^3x$. But in YM, the nonlinear YM charge density alters the constraint: the matter current alone is NOT conserved and even the total local current is not: the system requires a whole new concept --- that of asymptotic color Killing vectors $X^a$, $D_\mu X^a=0$ (or equivalently, Killing with respect to a suitable background field [4]) to ``bleach" the color current and allow ordinary conservation, of the ``$a$" direction (conserved) charge. This in turn means that it is the boundary conditions that determine when there are conserved integrated (only) charges, NOT just the field equations: local gauge invariance must be broken. This spells out why Noether and Hilbert's covariant attempts failed.

\section{Gravity}
The basic message is that gravity weighs, just as YM has color. For concreteness, we work with the original GR Einstein-Hilbert action; nothing essential changes in more elaborate models, including its cosmological constant [5], or higher derivative, extensions [6]. [The original ADM treatment [3] defined energy and the other constants of motion, without explicit use of the more elaborate explicit asymptotic Killing vector language of [5,6].]

Einstein theory's "left-hand sides", $G^{\beta\mu}$, obey the contracted Bianchi identities, $D_\beta G^{\beta\mu}=0$.
Separately, the covariantized --- in an arbitrary metric --- matter sources $T^{\beta\mu}$ are also (covariantly) conserved by virtue of their own field equations. Clearly, only a vector (density) current can be ordinarily conserved, $\dd_\mu j^\mu=0$, and so produce an integrated charge $ \int d^3x j^0$; a tensor one cannot, a hint that $T^{\mu\beta}$ is not (quite) the right current. To reduce this tensor to a vector, the only possibility is contraction with a vector $X_\beta$. But the resulting vector is no longer conserved, unless $X$ satisfies the Killing equations $D_\beta X_\mu+ D_\mu X_\beta=0$. On the other hand, the existence of such vectors is too strong a constraint on allowed spacetime solutions of Einstein's (or its generalizations') equations. Fortunately, a much weaker --- and physically quite satisfactory --- requirement is involved: it suffices for the Killing vectors to exist at spatial infinity. This requires space to be asymptotically flat/(A)dS (at a well-defined rate), so that it admits the Poincare/(A)dS, group there. The boundary conditions break the coordinate invariance obstruction down to a constant parameter group, but only at the edges. The physics is also perfect: one would hardly expect energy to be well-defined if the excitations did not subside outside a finite region or if space is closed! Also note how the specific asymptotic character affects the constants of motion: Poincare and (A)dS are very different groups. One may wonder how asymptotic conditions suffice to ``integrate" a volume charge; the answer is simply that the constraint is always a Poisson equation, linear on its left side, whose volume integral reduces to a surface integration at the (infinite) boundary --- one does not evaluate the total charge in the interior; rather, it is read off from the asymptotic field, just as for Maxwell's $V\sim (Q= \int d^3x j^0)/r$, $\dd_0 Q=0$.

Another totally equivalent way to understand symmetry breaking in the 
gravitational context is as the introduction of a preferred 
background geometry into models, like GR, that are precisely built 
not to depend on any. Once introduced, these geometries provide the 
desired global -- by breaking the local -- symmetries for the theory expanded about them. 
That is, only those solutions of (say) GR that  vanish at spatial 
infinity are now relevant. This “background field method” was used in [5] and implicitly by [3].

\section{Envoi}
We have attempted to show that all is for the best in the best of all possible YM and gravitational worlds, by reconciling Noether's two theorems with the physics of these systems, through symmetry breaking by boundary conditions. The idea can also be understood from another, seemingly different, aspect of all generally covariant models --- their ``already parametrized", Jacobi, form of the action principle,
\begin{equation}
I[\hbox{grav}] = \int d^n x \left[ \pi^{i j} \dd_0 g_{ij} - N_\mu R^{\mu}(\pi,g)\right], \hbox{ $i,j = 1 \ldots n-1$,}
\end{equation}
the basis of the entire ADM analysis. Here $g_{ij}$ are the spatial metric's covariant components and $\pi^{ij}$ their conjugate momenta (whose values are obtained by varying $\pi$) --- essentially $\dd_0 g_{ij}$. The $N_\mu$ are $n$ Lagrange multipliers (essentially the $g_{0\mu}$) that enforce the n constraints $R^\mu=0$ --- so the ``Hamiltonian" seems to vanish --- as it must, since coordinate invariance does not provide a preferred time direction. The invariant matter actions also have the ``Jacobi" form, 
\begin{equation}
I[\hbox{matt}] \sim  \int \left[ p^a \dd_0 q_a - N_\mu R^\mu(p, q, g_{ij}, \pi^{ij})\right].    
\end{equation}
The moral is the same: Only when a (not too crazy) choice of time is made, i.e., when diffeomorphism invariance is broken to constant Poincare or (A)dS at infinity, can the concepts of Hamiltonian and energy emerge.

\subsubsection*{Acknowledgements}
This work was supported by the U.S. Department of Energy, Office of Science, Office of High Energy Physics, under Award Number de-sc0011632. I thank A. Schwimmer for encouragement and J. Franklin for tech.

\end{document}